%% file: paper.tex
\documentclass[journal]{IEEEtran}
\IEEEoverridecommandlockouts
\usepackage[noadjust]{cite}
\usepackage{amsmath}
\usepackage[mathscr]{euscript}
\usepackage{algorithmic}
\usepackage{array}
\usepackage{mdwmath}
\usepackage{amssymb}
\usepackage{mathtools}
\usepackage{eqparbox}
\usepackage{mdframed}
\usepackage{steinmetz}
\usepackage{stfloats}
\usepackage{graphicx}
\usepackage{url}
\usepackage{tikz}
\usepackage{pgfplots}
\usepackage{subcaption}
\graphicspath{{../pdf/}{../jpeg/}}
\DeclareGraphicsExtensions{.pdf,.jpeg,.png}
\usepackage{epstopdf}
\usepackage{amsfonts}
\begin{document}
\title{Intelligent Reflecting Surface Assisted Wireless Communication: Modeling and Channel Estimation}
\author{Qurrat-Ul-Ain~Nadeem,~\IEEEmembership{Member,~IEEE,} Abla~Kammoun,~\IEEEmembership{Member,~IEEE,} Anas~Chaaban,~\IEEEmembership{Member,~IEEE,}
        M{\'e}rouane~Debbah,~\IEEEmembership{Fellow,~IEEE,} and~Mohamed-Slim~Alouini,~\IEEEmembership{Fellow,~IEEE}% <-this % stops a space
				\thanks{Q.-U.-A. Nadeem and A. Chaaban are with School of Engineering, The University of British Columbia, Kelowna, Canada (email: \{qurrat.nadeem, anas.chaaban\}@ubc.ca)}
\thanks{A. Kammoun and M.-S. Alouini are with the Computer, Electrical and Mathematical Sciences and Engineering (CEMSE) Division, King Abdullah University of Science and Technology (KAUST), Thuwal, Saudi Arabia 23955-6900 (e-mail: \{abla.kammoun,slim.alouini\}@kaust.edu.sa)}% <-this % stops a space
\thanks{M. Debbah is  with  CentraleSup{\'e}lec, Gif-sur-Yvette, France (e-mail: merouane.debbah@centralesupelec.fr).}% <-this % stops a space
}
\maketitle
\begin{abstract}
The recently completed 5G new radio standard is a result of several cutting-edge technologies, including massive multiple-input multiple-output (MIMO), millimeter (mm)-Wave communication and network densification. However, these technologies face two main practical limitations 1) the lack of control over the wireless channel, and 2) the high power consumption of the wireless interface. To address the need for green and sustainable future cellular networks, the concept of reconfiguring wireless propagation environments using Intelligent Reflecting Surfaces (IRS)s has emerged. An IRS comprises of a large number of low-cost  passive antennas that can smartly reflect the impinging electromagnetic waves for performance enhancement. This paper looks at the evolution of the reflective radio concept towards IRSs, outlines the IRS-assisted multi-user multiple-input single-output (MISO) communication model and discusses how it differentiates from the conventional multi-antenna communication models. We propose a minimum mean squared error (MMSE) based channel estimation protocol for the design and analysis of IRS-assisted systems. Performance evaluation results at $2.5$ GHz operating frequency are provided to illustrate the efficiency of the proposed system.

\end{abstract}

%\begin{IEEEkeywords}
%\end{IEEEkeywords}
\vspace{-.14in}
\section{Introduction}

The Fifth Generation (5G) mobile communication standard promises to provide enhanced mobile broadband, massive connectivity and ultra-low latency through various technological advances, including massive multiple-input multiple-output (MIMO), millimeter wave (mmWave) communications, and network densification. However, these technologies  consume a lot of power  and struggle to provide the users with guaranteed quality of service (QoS) in harsh propagation environments. For example: the network's total energy consumption scales linearly with the numbers of base stations (BS)s and the active antennas at each BS, while  communication at mmWave bands suffers from high path/penetration losses. These limitations have resulted in the need for green and sustainable future cellular networks with  control over the propagation environment.

An emerging concept that addresses this need is that of a smart radio environment (SRE), where the wireless propagation environment is turned into an intelligent reconfigurable space that plays an active role in transferring radio signals from the transmitter to the receiver \cite{SRE}. This concept can be realized by deploying arrays of low-cost antennas \cite{ant}, smart reflect-arrays \cite{RA} and  reconfigurable meta-surfaces  \cite{SRE}  in the environment, to shape the impinging electromagnetic (EM) waves in desired ways in a passive manner, without generating new radio signals and thereby without incurring any additional power consumption. SRE is a recent but feasible concept with a lot of current research focusing on fabricating relevant hardware prototypes, implementing testbeds and doing system-level simulations \cite{SRE, LIS_mag1, LIS4}.

 Very recently, works approaching this subject from the wireless communication design perspective have appeared, in which the SRE is enabled by intelligent reflecting surfaces (IRSs). The IRS is viewed as a planar array of a large number of passive reflecting antennas. Joint designs for precoding at the BS and reflect beamforming at the IRS are proposed to achieve different communication goals, for example: improve the network's spectral or energy efficiency, increase the secrecy capacity for physical layer security, assist in wireless power transfer \cite{8741198 , LIS_jour, LISA}. Almost all existing  works assume the IRS to have perfect channel state information (CSI), which is highly unlikely in practice,  given the IRS has no radio resources of its own to estimate the channels.  

%In fact, channel estimation is one of the biggest challenges in the practical design of IRS-assisted communication systems.

%Moreover, these preliminary works consider Rayleigh fading channel coefficients to model the communication links, which is an oversimplification of the conditions encountered in realistic propagation environments. In fact, the choice of channel model can significantly impact the resulting performance prediction. This is an important contribution as channel estimation has been repeatedly identified as a big challenge to the practical implementation of these systems and yet ignored in recent works on IRS-assisted communication systems.

In this paper, we outline the IRS-assisted system model and discuss how it differentiates from conventional multi-antenna communication systems.  We first discuss the evolution of the reflective radio concept towards IRSs along with its implementations and applications in wireless communications in Section II. The communication model for an IRS-assisted MISO system is introduced in Section III, and comparisons are made with existing MISO communication models.  We propose a novel minimum mean squared error (MMSE) based channel estimation protocol in Section IV by forming a control loop between the BS and IRS to serially estimate each IRS-assisted link.  Simulation results in Section V show the IRS-assisted system to be highly efficient but also sensitive to CSI errors.   Some related research directions and concluding remarks are presented in Section VI and VII respectively. 

\vspace{-.12in}

\section{Evolution towards IRSs}

Reflective radio technology has emerged as an attractive solution for designing energy and spectral efficient communication systems \cite{LISA}. In fact, the reflective devices (RDs), which do not use expensive and power-hungry active components, have become popular candidates to transmit signals to their receivers or improve the transmission of primary communication systems, using EM scattering of the radio waves. The former is known as backscatter communication (BSC) \cite{BS} while the latter is referred to as reflective relay \cite{LISA}. 

%For example, to transmit `$0$' and `$1$', the RD only needs two reflection states and the receiver can extract the information from the backscattered signal by recovering these reflection states.The deployment cost and radio spectrum usage of such BSC can become unaffordable with the massive number of IoT devices. To overcome this

Traditional BSC is widely used to support low power internet-of-things (IoT) communication, where the RD exploits its different reflection states to modulate its own messages over a continuous wave signal generated by a dedicated radio frequency (RF) emitter. A basic RD circuit varies the load impedance connected to its antenna through a switch to realize  different  reflection states, e.g. different phase values. The more recent  ambient BSC exploits RF signals transmitted from an existing primary source like the cellular BS or WiFi access point and therefore no dedicated RF emitter is needed.

%, reducing the deployment cost and radio spectrum usage. 

Reflective relays have been used to improve the QoS of blocked or weak users in the primary system. Conventionally, a single antenna is deployed in the reflective relay, leading to a weak reflective link, which can be improved by using reflective arrays instead. The concept of reflective array was first proposed in \cite{rff}, when open-ended wave-guides were used as antenna elements to change the phase of the reflected signal. Interest in reflective arrays increased in 1990s after micro-strip patch antennas were used to implement them and found applications in radar and satellite communication \cite{pat}. Although there have been conceptual discussions on reconfigurable reflective arrays, but hardware prototypes have been made available only recently \cite{SRE}. This has led to the concept of an IRS, which is a 2D structure of a large number of passive elements/antennas with the ability of reconfiguring the incident EM waves in real time. The first works that consider the use of IRSs to re-program indoor and outdoor propagation environments appeared in 2012  \cite{LIS4} and 2018 \cite{8741198} respectively.

 %The traditional reflective arrays have had variety of applications in radar and satellite communications but their use in terrestrial wireless communication systems was not considered earlier 
%as they could not adapt the induced phases with the time-varying channels that constitute the
%wireless propagation environments.
\begin{figure}
\centering
\includegraphics[width=2.7in, height=1.5in]{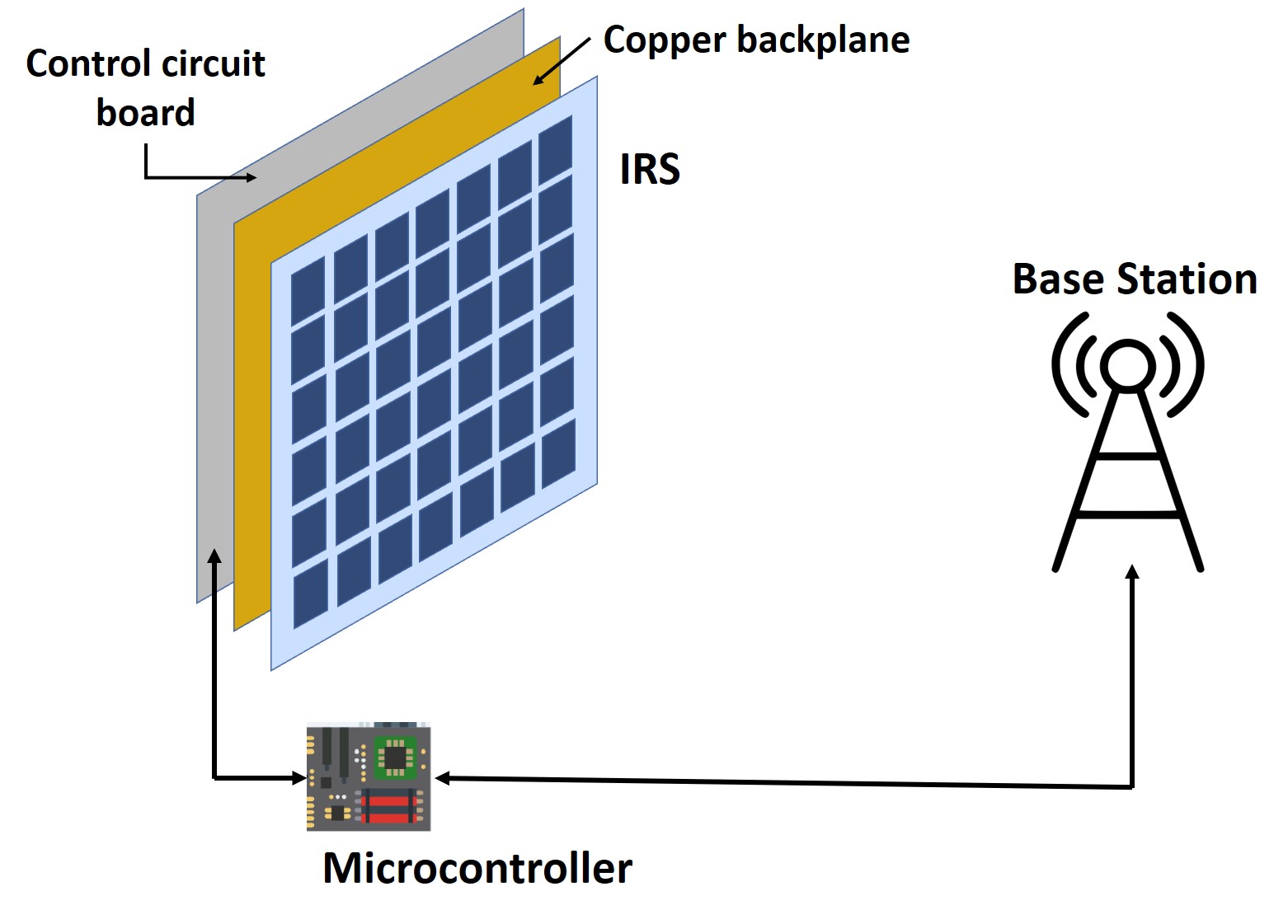}
\caption{Typical implementation of an IRS system.}
\label{IRSsys}
\end{figure}

We will now briefly discuss some current implementations of IRSs. The authors in \cite{ant} present the implementation directions for embedding arrays of low-cost antennas into the walls of a building to passively reflect the incident wireless signals. The developed prototype uses three parabolic antennas loaded with single-pole four-throw switches and is able to achieve $64$ different reflection configurations.  In \cite{RA}, a reflect-array  with $224$ reflecting units is fabricated, with each unit formed by loading a micro-strip patch element with an electronically-controlled relay switch. To continuously tune the phase responses of the reflecting units, variable capacitors are integrated into the reflector panel.  Alternate designs include using  positive-intrinsic-negative (PIN) diodes, field-effect transistors (FETs), or micro-electromechanical system (MEMS) switches within the reflecting panel. Current reflect-array designs allow the configuration of the reflecting elements at the code-book level, such that the whole surface can be electronically shaped to adaptively synthesize different beam shapes. Moreover,  the design of reconfigurable meta-surfaces, composed of thin meta-material layers of a large number of sub-wavelength scattering particles (meta-atoms), has provided the capability  of shaping the radio waves in fully customizable ways with properties like negative refraction, anomalous reflection and perfect absorption \cite{SRE}. The communication model in this work assumes the IRS to be realized using an array of antennas that passively reflect the incoming waves.

Fig. \ref{IRSsys} shows a typical implementation of an IRS system which consists of the IRS realized using patch antennas printed  on  a dielectric  substrate, a copper back-plane to prevent signal leakage and a control circuit board that adjusts the reflection amplitude and phases of the IRS elements, as triggered by a small micro-controller. In practice, field-programmable gate array (FPGA) can  be  implemented  as  the  microcontroller,  which  also  acts  as  a gateway  to  communicate/coordinate  with the BS (where all beam control is implemented) through a separate wired or wireless backhaul link.

%Although, the traditional reflective arrays have had variety of applications in radar and satellite communications but their use in terrestrial wireless communication systems was not considered earlier as they could not adapt the induced phases with the time-varying channels that constitute the wireless propagation environments. It is the availability of reconfigurable reflective arrays that led to the concept of IRS-assisted wireless communication systems. 

The most attractive application of IRS is to act as a reflective relay to improve the QoS of users suffering from unfavorable propagation conditions. In this case, it resembles a full-duplex (FD) multi-antenna amplify-and-forward (AF) relay. However,  the FD-AF relay needs active electronic components, such as digital-to-analog convertors (DACs), analog-to-digital converters (ADCs), power amplifiers, as well as self-interference cancellation circuits.  In contrast, IRSs are meant to be realized with minimal hardware complexity and power requirements.  Moreover, the received SNR through the IRS-assisted link is shown to scale quadratically in the number of reflecting elements \cite{LISA} as opposed to the classical beamforming methods at the BS and AF relays, where the SNR scales linearly with the number of antennas. With a large number of elements, IRS can be much more advantageous than AF relays and has recently found applications in physical layer security as well as simultaneous wireless information and power transfer systems.

%The system model for AF relay-assisted and IRS-assisted communication also have some key differences, as will be highlighted in the next section.

Another recent application of IRSs is in mmWave communication systems, where a BS comprising of few active antennas illuminates a nearby large IRS \cite{mm23}. By scaling up the number of passive elements at the IRS without increasing the number of active antennas at the BS, massive MIMO beamforming gains are yielded.  However, these work assume a loss-less fixed connection between the BS and the IRS by placing the latter very close to the BS. Different from this, we consider the use of IRSs in the propagation environment to enhance coverage and QoS, with the channel between the BS and IRS not being loss-less and not necessarily being fixed. In the next section, we study the IRS-assisted system model along with how it differentiates from conventional MISO, hybrid MISO as well as relay-assisted communication models. 

%We also tackle the CSI acquisition challenge associated with IRSs by  proposing a practical channel estimation protocol to enable precoding at the BS and reflect beamforming at the IRS. 

%In fact, our system model allows the use of IRSs for several other applications like to spectrum sharing in cognitive radio neworks, increase secrecy capacity in terms of physical layer security as well as act as assist wireless power transfer. 

%In the next section, we present the system model for IRS-assisted downlink transmission and discuss how  the constraints the IRS elements have to meet differ from those of conventional MIMO architectures. 

%Alternative designs include the use of varactor diodes and  MEMS, where the reflector units can be electronically controlled to change the resonant frequency and create desired phase and amplitude change \cite{}. 
%
%
 %Even though in the proposed reflect-array in \cite{}, the phases induced by the reflector units are controller individually, but with significant progress made in the design and fabrication of reflect-arrays, the phase control at individual unit level is not needed. Rather, the reflecting elements can be configured at the codebook level and  the   phase   coefficients can be optimized jointly to obtain   the   desired   radiation   pattern and beam shapes.   

\vspace{-.1in}
\section{Communication Model}

%\begin{figure*}[t!]
%\begin{mdframed}[linewidth=2pt]
%\begin{center}
%\subfigure[IRS-assisted multi-user MISO system. Red dashed lines represent the estimated uplink channel vectors.]{
            %\label{SU1}
            %\includegraphics[width=0.45\textwidth]{LIS_layout.jpg}
        %}
        %\subfigure[Channel estimation protocol.]{
           %\label{SU2}
           %\includegraphics[width=0.51\textwidth]{CE.jpg}
        %}
%%\label{singleuserlayout}
%\end{center}
%\caption{Proposed IRS-assisted multi-user MISO communication model and channel estimation protocol.}
   %\label{SU}
	%\end{mdframed}
%\end{figure*}

\begin{figure*}[!t]
\begin{subfigure}[t]{.48\textwidth}
\includegraphics[scale=.25]{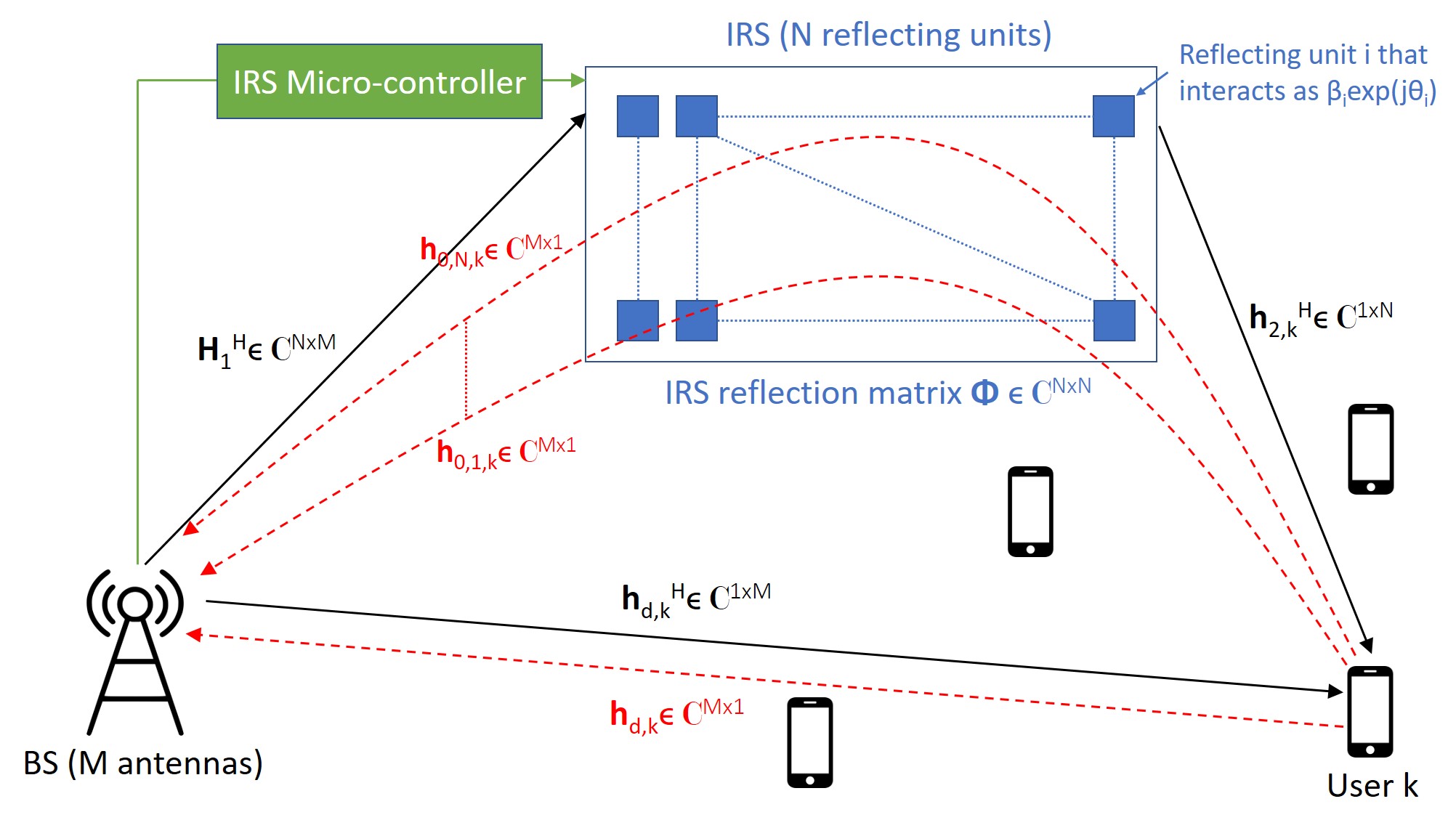}
\caption{IRS-assisted multi-user MISO system. Red dashed lines represent the estimated uplink channel vectors.}
\label{SU1}
\end{subfigure}
\hspace{.05cm}
\begin{subfigure}[t]{.48\textwidth}
\includegraphics[scale=.28]{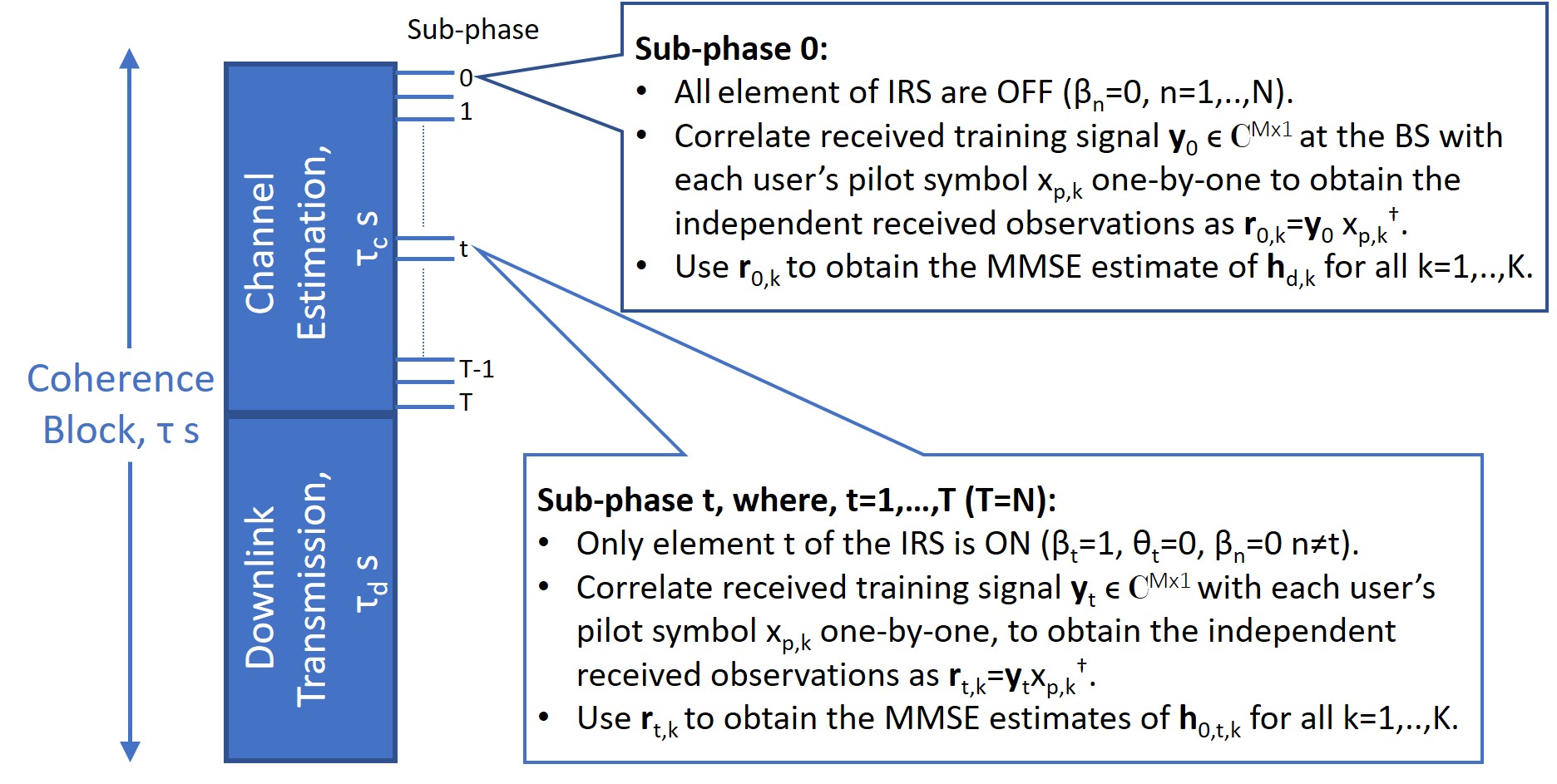}
\caption{Channel estimation protocol.}
\label{SU2}
\end{subfigure}
\caption{Proposed IRS-assisted multi-user MISO communication model and channel estimation protocol.}
\label{SU}
\end{figure*}

The proposed IRS-assisted multi-user MISO system is illustrated in Fig. \ref{SU1}, which consists of a BS equipped with $M$ antennas serving $K$ single-antenna  users. This communication is assisted by an IRS, comprising of $N$ nearly passive antennas which introduce phase shifts onto the incoming signal waves, attached to the facade of a building in the propagation environment. As discussed already in Fig. \ref{IRSsys}, the reflection configuration of the IRS (i.e. the phase shifts applied by individual IRS elements) is controlled by a micro-controller, which gets this information from the BS over a  backhaul link. 

%For example: the information can be provided in the form of an index of a pre-defined quantized codebook shared by the BS and micro-controller, that contains information on the various reflection configurations possible at the IRS.

The BS transmits the signal vector $\mathbf{x}=\sum_{k=1}^K \sqrt{p_k}\mathbf{g}_k s_k $, where $p_k$, $\mathbf{g}_k\in \mathbb{C}^{M \times 1}$ and $s_k$ are the allocated power, digital precoding vector and data symbol of user $k$ respectively. Given $s_k$'s are i.i.d. $\mathcal{CN}(0,1)$ variables, $\mathbf{x}$ has to satisfy the power constraint $\mathbb{E}[||\mathbf{x}||^2]=\text{tr}(\mathbf{P}\mathbf{G}^H\mathbf{G})=P_{T}$, where $\textbf{G}=[\mathbf{g}_1, \dots, \mathbf{g}_K]\in \mathbb{C}^{M\times K}$ and $\mathbf{P}=\text{diag}(p_1, \dots, p_K)\in \mathbb{C}^{K\times K}$. The received signal at user $k$ is given as, \vspace{-.05in}
\begin{align}
y_{k}&=(\mathbf{h}_{d,k}^H+\mathbf{h}_{2,k}^H \boldsymbol{\Phi}^H \mathbf{H}_{1}^H)\mathbf{x}+n_{k}, \nonumber \\
\label{ch1}
&=(\mathbf{h}_{d,k}^H+\mathbf{v}^H \mathbf{H}_{0,k}^H)\mathbf{x}+n_{k}
\end{align}
where $\mathbf{H}_{1}^H \in \mathbb{C}^{N\times M}$ is  the channel from the BS to the IRS, $\mathbf{h}_{2,k}^H \in \mathbb{C}^{1\times N}$ is the channel from the IRS to user $k$, $\mathbf{h}_{d,k}^H \in \mathbb{C}^{1\times M}$ the direct channel from the BS to user $k$ and $n_{k}\sim \mathcal{CN}(0,\sigma_n^2)$ is the noise at the user. The IRS is represented by the diagonal matrix $\boldsymbol{\Phi}=\text{diag}(\beta_1 \exp(j\theta_{1}), \dots, \beta_N \exp(j\theta_{N}))$, where $\theta_{n} \in [0,2\pi ]$ and $\beta_{n} \in [0,1]$ represent the phase and the amplitude coefficient for element $n$ respectively. The uplink channel through the IRS $\mathbf{H}_{1} \boldsymbol{\Phi} \mathbf{h}_{2,k}$ can be equivalently expressed as $\mathbf{H}_{0,k} \mathbf{v}$, where $\mathbf{v}=[\beta_1 \exp(j\theta_{1}), \dots, \beta_N \exp(j\theta_{N})]^T \in \mathbb{C}^{N\times 1}$ and $\textbf{H}_{0,k}=\textbf{H}_1\text{diag}(\textbf{h}_{2,k}^{T})\in \mathbb{C}^{M\times N}$. This formulation separates the  response of the IRS in $\mathbf{v}$ from the cascaded channel outside the IRS control in  $\mathbf{H}_{0,k}$, and will assist us in the design of the channel estimation protocol in Section IV. 
%In practice, it is very costly to achieve continuous phases over the $[0,2\pi ]$, so $\theta_n$s usually take only discrete values as $\theta_n \in \{0, \frac{2\pi}{Q}, \dots, \frac{2\pi(Q-1)}{Q}  \}$, where $Q$ is the number of possible quantized phase values. 

We will now discuss how this model is different and more challenging to tackle in terms of beamforming design and CSI acquisition when compared to other multi-antenna models.

\textit{Conventional MISO Communication Model:} Under this model, the received signal at user $k$ is given as $y_{k}=\mathbf{h}_{d,k}^H\mathbf{x}+n_{k}$. Finding the optimal $\mathbf{G}$ and $\mathbf{P}$ that maximize the optimization criteria of interest, like sum-rate, minimum rate, energy efficiency, subject to $\text{tr}(\mathbf{P}\mathbf{G}^H\mathbf{G})=P_{T}$, are well-studied problems in literature with optimal solutions, e.g. dirty paper coding for sum-rate, optimal linear precoder in \cite{LIS_jour} for minimum rate. The channels are estimated at the BS either by exploiting channel reciprocity in time division duplexing (TDD) systems or by explicit downlink training  and CSI feedback in frequency division duplexing (FDD) systems \cite{massiveMIMOO}.

\textit{Relay-Assisted MISO Communication Model:} The received signal at user $k$ is given as  $y_k= \mathbf{h}_{2,k}^H \mathbf{V}^H \mathbf{y}_R + \mathbf{h}_{d,k}^H \mathbf{x}+ n_k$, where $\mathbf{y}_R=\mathbf{H}_1^H \mathbf{x} + \mathbf{n}_R$, $\mathbf{V}\in \mathbb{C}^{N\times N}$ is the diagonal AF matrix of the relay and $\mathbf{n}_R \sim \mathcal{CN}(\mathbf{0},\sigma^2 \mathbf{I}_N)$, where  $N$ is the number of antennas at the relay. The design goal is to find the optimal $\mathbf{P}$, $\mathbf{G}$ and $\mathbf{V}$ given $\text{tr}(\mathbf{P}\mathbf{G}^H\mathbf{G})=P_{T}$ and $\mathbb{E}[\text{tr}(\mathbf{V}^H\mathbf{y}_R\mathbf{y}_R^H \mathbf{V})]=P_{R}$, where $P_R$ is the Tx power at the relay. The BS estimates $\mathbf{H}_1$ using the pilot symbols sent by the relay while $\mathbf{h}_{2,k}$ is estimated at the relay based on the pilot symbols sent by the users and the CSI is shared between the relay and BS.  The joint design of $\mathbf{G}$ and $\mathbf{V}$ is challenging but several efficient algorithms exist in literature \cite{reln}. 

\textit{Hybrid Beamforming Model in mmWave Systems: }  To reduce the number of RF chains needed for fully digital precoding in mmWave massive MISO systems, hybrid beamforming architectures have been proposed where the overall beamformer consists of a low-dimension digital beamformer followed by the analogue beamformer \cite{CE_mm, mm23}. The received signal at user $k$ is given as $y_{k}=\mathbf{h}_{d,k}^H\mathbf{x}+n_{k}$. The Tx signal $\mathbf{x}=\sum_{k=1}^K \sqrt{p_k} \mathbf{V}_{A} \mathbf{g}_k s_k$, where $\mathbf{V}_A \in \mathbb{C}^{M\times M_{RF}}$ is the analogue beamforming matrix with constraints $|\mathbf{V}_A(i,j)|=1$, $i=1,\dots, M$, $j=1,\dots, M_{RF}$ and $\mathbf{g}_k \in \mathbb{C}^{M_{RF}\times 1}$ is the digital precoding vector, where $M_{RF}$ is the number of RF chains. The channels $\mathbf{h}_{d,k}$s are estimated at the BS under TDD and at the users (who then feedback this CSI to the BS) under FDD. Algorithms based on compressive sensing and channel dimension reduction, to reduce the huge CSI feedback overhead incurred in massive MISO systems,  have been proposed \cite{CE_mm}. The joint design of $\mathbf{V}_A$ and $\mathbf{G}$ is challenging because of the unit-modulus constraints on $\mathbf{V}_A$. This non-convex problem can not be be solved optimally using standard methods which has resulted in several suboptimal solutions.

In terms of precoding/beamforming design, the IRS-assisted system model is much more difficult to analyze as compared to the first two models, due to the unit-modulus constraints on elements of the reflect beamforming matrix $\boldsymbol{\Phi}$. Although beamforming optimization under unit-modulus constraints has been studied in the context of hybrid digital/analog mmWave architectures \cite{CE_mm, mm23}, such designs are mainly restricted to the BS  side, and are not directly applicable to the joint design of the precoding at the BS and reflect beamforming at the IRS. In terms of CSI acquisition, the IRS-assisted system is different from all of the above models as IRS has no radio resources of its own to send pilot symbols to help the BS estimate $\mathbf{H}_1$ nor can it receive and process pilot symbols from the users to estimate $\mathbf{h}_{2,k}$s.  This is one of the biggest challenges in the practical design of IRS-assisted systems \cite{LISA}.

The design of IRS-assisted systems also requires the correct modeling of $\mathbf{h}_{2,k}$ and $\mathbf{H}_{1}$. Existing works (e.g. \cite{8741198}) utilize the independent Rayleigh model, which is not practical unless the IRS elements are spaced far enough and the environment has rich scattering. Moreover, under this setting, the statistics of the received signal do not change with the values of the phase shifts as $N,M$ grow large so asymptotically there will be no significant performance improvement due to IRS. It is therefore important to analyze the performance under the more practical ray-tracing or correlated Rayleigh channel models.

The  IRS is envisioned to be installed on a high rise building which will likely result in a  LoS channel between the BS and the IRS \cite{LIS_jour}.   There are two ways to model this LoS channel. \\
\textbf{1) Rank-One Channel:} Since the BS and the IRS have co-located elements, so the channel matrix $\mathbf{H}_{1}$ will have rank one, i.e. $\mathbf{H}_{1}=\mathbf{a}\mathbf{b}^{H}$, where $\mathbf{a} \in \mathbb{C}^{M\times 1}$ and $\mathbf{b}\in\mathbb{C}^{N\times 1}$ describe the array responses at the BS and the IRS. The degrees of freedom offered by the IRS-assisted link will be one and the IRS will only yield performance gains when $K=1$ \cite{LIS_jour}.  \\
\textbf{2) Full-rank Channel:} To benefit from the IRS in the multi-user setting, we must have $\text{rank}(\mathbf{H}_{1})\geq K$. One way to introduce this rank is to have scattering between the BS and the IRS. Also using distributed IRSs, the LoS channel matrix between the BS and the IRSs can be made of high rank.

%In addition to channel modeling, the design of channel estimation protocols to obtain channel state information (CSI) poses another main challenge to the design of IRS-assisted communication systems, because the IRS has no active resources of its own to sense the channel and relies completely on the IRS-controller for its operation. It is due to this difficulty, existing works on this technology assume perfect CSI to be available at both BS and IRS. In the next section, we propose an MMSE based channel estimation protocol  that does not require any active participation from the IRS. 
\vspace{-.07in}

 \section{Channel Estimation Protocol}

Channel estimation is necessary to compute the precoder $\mathbf{G}$ and the reflect beamforming matrix $\boldsymbol{\Phi}$. We focus on TDD systems and exploit channel reciprocity in estimating the downlink channels using the received uplink pilot signals from the users.  The channel coherence period of $\tau$s is divided into an uplink training phase of $\tau_{c}$s and a downlink transmission phase of $\tau_{d}$s. Throughout the training phase, the users transmit mutually orthogonal pilot symbols $x_{p,k}$, where $|x_{p,k}|^2=p_{c}$. 
%, to allow the BS to compute the estimates $\hat{\textbf{H}}_{1}$, $\hat{\textbf{h}}_{2,k}$s and $\hat{\textbf{h}}_{d,k}$s

The real difficulty is in the estimation of $\mathbf{H}_{1}$ and $\mathbf{h}_{2,k}$s as the IRS has no radio resources and signal processing capability to send or receive pilot symbols. Therefore, the BS has to estimate all the channels and use them to compute  $\mathbf{G}$ as well as $\boldsymbol{\Phi}$.  The BS then provides information on the required IRS reflection configuration (i.e. $\boldsymbol{\Phi}$ or $\mathbf{v}$ in \eqref{ch1}) to the micro-controller connected to the IRS. The information can be provided in the form of an index of a pre-defined quantized codebook shared by the BS and micro-controller, that contains the various reflection configurations possible at the IRS. Wireless backhaul links in the mmWave and tera-hertz bands are suitable candidates to enable the BS to communicate with the IRS controller under strict latency requirements \cite{SRE}.

%For example: the information 

To this end, note that $\mathbf{H}_{1}$ and $\mathbf{h}_{2,k}$ have been cascaded as $\mathbf{H}_{0,k}\in \mathbb{C}^{M\times N}$ in (\ref{ch1}), where $\mathbf{H}_{0,k}=[\mathbf{h}_{0,1,k}, \dots, \mathbf{h}_{0,N,K}]$ is a matrix of $N$ column vectors. Each vector $\mathbf{h}_{0,n,k} \in \mathbb{C}^{M\times 1}$ (shown in red curved arrows in Fig. \ref{SU1}) can be interpreted as the channel from the user to the BS through the IRS when only element $n$ of the IRS is ON i.e. $\beta_{n}=1, \theta_{n}=0$ and $\beta_{i}=0$, $i\neq n$. We will focus on the MMSE estimation of $\mathbf{h}_{0,n,k}$, $n=1,\dots, N$ and $\mathbf{h}_{d,k}$ for $k=1,\dots, K$ at the BS. 

%The total channel estimation time $\tau_{c}$ is divided into $T+1$ sub-phases, each of duration $\tau_{s}=\frac{\tau_{c}}{T+1}$s, where $T=N$.  During the first sub-phase, BS requests the micro-controller to keep all the IRS elements in the OFF state (i.e. $\beta_n=0$, $n=1,\dots, N$) and the BS estimates the direct channel $\textbf{h}_{d,k}$ for all users.  During the $(t+1)^{th}$ sub-phase, where $t=1,\dots, T$, the BS requests the micro-controller to switch element $t$ of the IRS ON (i.e. $\beta_{t}=1$, $\theta_{t}=0$)  to aid the BS in estimating $\textbf{h}_{0,t,k}$ for all users, while all other IRS elements are kept OFF.

The total channel estimation time $\tau_{c}$ is divided into $T+1$ sub-phases, each of duration $\tau_{s}=\frac{\tau_{c}}{T+1}$s, where $T=N$.  During sub-phase $0$, the BS requests the micro-controller to keep all the IRS elements in the OFF state (i.e. $\beta_n=0$, $n=1,\dots, N$) and the BS estimates the direct channel $\mathbf{h}_{d,k}$ for all users.  During the $t^{th}$ sub-phase, where $t=1,\dots, T$, the BS sends a signal to the micro-controller to switch element $t$ of the IRS ON (i.e. $\beta_{t}=1$, $\theta_{t}=0$) while keeping all other elements OFF so that the BS can estimate $\mathbf{h}_{0,t,k}$. The micro-controller therefore triggers the control circuit board of the IRS in Fig. \ref{IRSsys} to implement the required $\mathbf{v}$, e.g., $\mathbf{v}=[1, \dots, 0]$ in sub-phase $1$, $\mathbf{v}=[0,1,0,\dots, 0]$ in sub-phase $2$ and so on. 

%such that the $i^{th}$ column, $i=1,\dots, N$, has only the $i^{th}$ element set as one while the others are set as zero. Also, the $(N+1)^{th}$ column has all zeros the $N\times (N+1)$ IRS phase shift matrix can be defined as $\bar{\boldsymbol{\Phi}}=[\textbf{0}_{N} \textbf{I}_{N}]$, where $\textbf{0}_{N}$ is an $N\times 1$ all-zero vector. 

 %For the channel estimation phase of $(N+1)\tau_{s}$ s, the received training signal matrix at the BS can be defined as $\textbf{Y}=[\textbf{y}_{1}, \textbf{y}_{2}, \dots, \textbf{y}_{N}, \textbf{y}_{N+1}] \in \mathbb{C}^{M\times N+1}$, where

The expressions of the MMSE estimates for all the channel vectors can be derived straightforwardly as follows. Let $\mathbf{y}_{0}$ represent the received training signal vector when all the elements of the IRS are OFF, given as,
\begin{align}
&\mathbf{y}_{0}=\sum_{k=1}^{K}\mathbf{h}_{d,k}x_{p,k}+\mathbf{w}_{B},
\end{align}
where $\mathbf{w}_{B}\sim \mathcal{CN}(\textbf{0},\sigma^2\textbf{I}_{M}) \in \mathbb{C}^{M\times 1}$ is the noise vector at the BS.  The BS uses $\mathbf{y}_{0}$ to estimate $\mathbf{h}_{d,k}$, $\forall k$. Similarly,  $\mathbf{y}_{t}, t=1,\dots, T$, represents the received training signal vector when  element $t$ of the IRS is ON and is used to estimate $\mathbf{h}_{0,t,k}$ for all the users. It can be expressed as,
\begin{align}
&\mathbf{y}_{t}=\sum_{k=1}^{K}\mathbf{h}_{d,k}x_{p,k}+\mathbf{h}_{0,t,k}x_{p,k}+\mathbf{w}_{B}, \hspace{.04in} t=1,\dots, T.
\end{align}

Note that each $\mathbf{y}_{i}$, $i=0,1,\dots, T$, contains the received signals from all the $K$ users. However since the users transmit mutually orthogonal pilot symbols, so the received training signal vector $\mathbf{y}_{i}$ is correlated one-by-one with the pilot symbol of each user to obtain the observation vector with respect to each user as $\mathbf{r}_{i,k}=\mathbf{y}_{i}x_{p,k}^{\dagger}$, where $\dagger$ represents the pseudo-inverse. These independent observation vectors are then used to obtain the estimates of the channel vector being estimated in the $i^{th}$ sub-phase for all the users.  The estimates are computed using the MMSE estimation method \cite{massiveMIMOO}.  This protocol is also summarized in Fig. \ref{SU2}. The BS then uses the channel estimates to compute the optimal beamforming vector $\mathbf{v}^*$ in (\ref{ch1}) and sends this information to the IRS microcontroller.

%We will show in the next section that channel estimation in an IRS-assisted system is more prone to errors than in a conventional MISO communication system. 

%
%\section{IRS Phase Matrix Design}
%
%The design of the IRS phase matrix for the transmission phase largely depends on the performance criteria being optimized. Since in practice, each element on the IRS is designed to maximize signal reflection so $\beta_{i} = 1$, $\forall i$. However, any design must satisfy the unit-modular constraints on the IRS elements, i.e. $|[\boldsymbol{\Theta}]_{i,i}|=1$. These non-convex constraints can be generally handled using projection methods to obtain the optimal solutions \cite{}. 
%
%The existing preliminary designs for the IRS phase shift matrix can be classified broadly as either aiming to maximize the network spectral efficiency or the network energy efficiency. An important work that deals with the former is \cite{LIS_jour}, where the authors propose a joint design for the precoding vectors at the BS and the IRS phase matrix that maximize the minimum user rate. To do so, the authors utilize the optimal linear precoder (OLP), that maximizes the minimum SINR subject to a given power constraint for any given IRS phase matrix, develop accurate deterministic approximations for the parameters of the asymptotically OLP, which are then utilized to optimize the LIS phase matrix based on the projected gradient descent method. An important contribution that deals with the maximization of the network energy efficiency appears in \cite{} where the authors  
\vspace{-.07in}
\section{IRS Design and Evaluation Results}

Assuming that the IRS elements have perfect reflection coefficient during downlink transmission, i.e. $\beta_{n} = 1$, for $n=1,\dots, N$, the IRS reflect beamforming matrix $\boldsymbol{\Phi}$ must satisfy the unit-modulus constraints on its diagonal elements, i.e. $|[\boldsymbol{\Phi}]_{n,n}|=1$, for $n=1,\dots, N$. For the single-user system, the vector $\mathbf{v}=[\exp(j\theta_{1}), \dots, \exp(j\theta_{N})]^T$ is designed so as  the reflected signals from the IRS add constructively with the ones received directly from the BS. Therefore the optimal $\boldsymbol{\Phi}^*=\text{diag}(\mathbf{v}^*)$, where $\mathbf{v}^*=\exp(j\phase{\mathbf{H}_{0}^H \mathbf{h}_{d}})$, where $\phase{\mathbf{x}}$ returns the vector of phases of $\mathbf{x}$. For the multi-user system, we focus on minimum user rate as the performance criteria and optimize the IRS phases using an adaptation of \textbf{Algorithm 1} from [7], which uses project gradient ascent to find $\boldsymbol{\Phi}^*$.

%Since this paper intends to be non-technical and provides a flavor of the performance gains realizable through the deployment of an IRS, so we utilize a basic design for the IRS phase shift matrix. The design referred to as Center of Means (CoM) in \cite{LIS_jour}, assigns the phase shifts $(\theta_{1}, \dots, \theta_{N})$ so as to reflect the incoming signals in the direction of the mean angle of arrival (AoA) of all the $K$ users.  

\begin{table}[!b]
\centering
\normalsize
\caption{Simulation parameters.}
\begin{tabular}{|l|l|}
\hline
  \textbf{Parameter} & \textbf{Value} \\ 
\hline
\textbf{Array parameters:} & \\
\hline
Carrier frequency & $2.5$ GHz \\
 BS, IRS configuration & Uniform linear array (ULA),\\
%& $0.5\lambda$ spacing\\
 %BS configuration & ULA, $0.5\lambda$ spacing \\
Tx power budget ($p_{T}$)& $5$ W \\
Noise level & $-80$\rm{dBm} \\
 \hline
\textbf{Path Loss:} & \\
\hline
Model & $\frac{10^{-C/10}}{d^{\alpha}}$ \\
$C$ (Fixed loss at $d=1$m)  & $26$\rm{dB} ($\textbf{H}_{1}$), $28$\rm{dB} ($\textbf{h}_{2,k},\textbf{h}_{d}$)\\
$\alpha$ (Path loss exponent) & $2.2$ ($\textbf{H}_{1}$), $3.67$ ($\textbf{h}_{2,k},\textbf{h}_{d}$)\\
\hline
\textbf{Channel Estimation:} & \\
\hline
 $\tau$, $\tau_{c}$, $\tau_{s}$ & $.01$s, $.01\tau$, $\frac{\tau_{c}}{N+1}$ \\
$p_c$ & $1$ W\\
\hline
\textbf{Channel Models} & \\
\hline
$\mathbf{H}_{1}$ in single-user case& Rank-One: $\mathbf{H}_{1}=\mathbf{a}\mathbf{b}^H$ \cite{LIS_jour} \\
$\mathbf{H}_{1}$ in multi-user case  & Full Rank \cite{LIS_jour}\\
$\mathbf{h}_{2,k}$ & Correlated Rayleigh: $\mathbf{R}_{IRS_{k}}^{1/2}\mathbf{z}_{k}$\\
$\mathbf{h}_{d,k}$ & Correlated Rayleigh:  $\mathbf{R}_{BS_{k}}^{1/2}\mathbf{z}_{d,k}$ \\
%$\textbf{z}_{k}$, $\textbf{z}_{d,k}$ & $\mathcal{CN}(\textbf{0}, \textbf{I}_{N})$\\
$\mathbf{R}_{BS_{k}}, \mathbf{R}_{IRS_{k}}$ & Generated using [\cite{LIS_jour} Sec. V] \\
\hline
\textbf{Precoding} & \\
\hline
Single-user case & Maximum ratio transmission \\
Multi-user case & Optimal Linear Precoder  \cite{LIS_jour} \\
\hline
\end{tabular}
\label{T1}
\end{table}

%\begin{figure*}[t!]
%\begin{mdframed}[linewidth=2pt]
%\begin{center}
%\subfigure[IRS-assisted single-user MISO system.  The BS and IRS are marked with their $(x,y,z)$ coordinates.]{
            %\label{SU1_sim}
            %\includegraphics[width=0.4\textwidth, height=.22\textwidth]{singleuserlayout.png}
        %} \hspace{.05in}
        %\subfigure[Performance of an IRS-assisted single-user MISO system  under perfect and imperfect CSI for $M=4$, $N=40$.]{
           %\label{SU2_sim}
           %\includegraphics[width=0.55\textwidth, height=.3\textwidth]{singleuserwithd.eps}
        %}
%\end{center}
%\caption{Results for a single-user system.}
   %\label{SU_sim}
	%\end{mdframed}
%\end{figure*}

\begin{figure*}[!t]
\begin{subfigure}[t]{.48\textwidth}
\includegraphics[scale=.6]{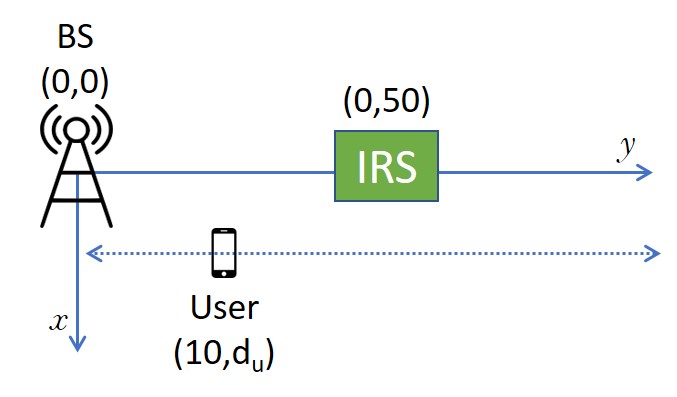}
\caption{IRS-assisted single-user MISO system.  The BS and IRS are marked with their $(x,y)$ coordinates.}
\label{SU1_sim}
\end{subfigure}
\hspace{.05cm}
\begin{subfigure}[t]{.48\textwidth}
\tikzset{every picture/.style={scale=.95}, every node/.style={scale=.8}}
\input{singleuserwithd.tex}
\caption{Performance of an IRS-assisted single-user MISO system  under perfect and imperfect CSI for $M=4$, $N=35$.}
\label{SU2_sim}
\end{subfigure}
\caption{Results for a single-user system.}
\label{SU_sim}
\end{figure*}

We utilize the parameter values described in Table \ref{T1}. The path loss values are computed  for $2.5$ GHz operating frequency using the 3GPP Urban Micro (UMi) parameters from TR36.814 (detailed in Section V of \cite{LIS_jour}). We use the LoS version to generate path loss for $\mathbf{H}_1$ and the non-LOS (NLOS) version to generate path losses for $\mathbf{h}_{2,k}$ and $\mathbf{h}_{d,k}$. Moreover, $5$ \rm{dBi} antennas are considered at the BS and IRS. Since IRS is deployed much higher than the BS for the purpose of avoiding the penetration losses that are caused by ground structures like buildings, we assume penetration losses of $10$ \rm{dB} for IRS-to-user link while that of $20$ \rm{dB} for BS-to-user link.     We first focus on the single-user scenario shown in Fig. \ref{SU1_sim} and study the received SNR in Fig. \ref{SU2_sim} by varying the value of $d_{u}$. We observe that in an IRS-assisted system, the user farther away from the BS can still be closer to the IRS and receive stronger reflected signals from it resulting in an improvement in the performance as observed for $d_{u}>30$. Consequently, the IRS-assisted system is able to provide a higher QoS to a larger region. For example, it will cover  $110$m with an SNR level of at least $5$\rm{dB}, whereas the system without the IRS can cover about $80$m to achieve the same SNR level. Moreover, the users placed close to the IRS, e.g. located in $42<d_u < 75$ range will see gains   ranging from $10$ to $17$ \rm{dB}.

%This extension is coverage is achieved using only a passive IRS that incurs negligible additional power consumption as compared to other solutions that increase coverage using an additional BS or active relay.  Although the SNR performance degrades due to increasing signal attenuation when $d_u>50$ but it is still better than what would have been achieved without the IRS unless the user is so far away that the path loss becomes dominant over the gain provided by the IRS.
% 
Doubling $N$ at the IRS to $70$, the received SNR scales by about $6$\rm{dB} for users close to the IRS, which implies the scaling to be in the order of $N^2$, corresponding to an array gain of $N$ and the reflect beamforming gain of $N$. However, the gain is negligible for $5<d_{u}<20$ because the BS-to-user direct channel is much stronger than the channel through the IRS. Moreover, much higher coverage is possible with large number of passive reflecting elements at the IRS.

%When the user is not close to either the BS or the IRS, the SNR gain is lower than $N^2$.Therefore, the performance of the IRS-assisted system deteriorates much more than the direct system as the  channel training SNR decreases ($\sigma^2$ increases). 

The result also shows that the IRS-assisted system is more sensitive to channel estimation errors than the conventional system. This is because for a constant channel estimation time of $\tau_{c}$s, the IRS-assisted system has to estimate $N+1=36$ channel vectors of dimension $M$ whereas the direct system only needs to estimate one channel vector. Moreover, the error becomes more significant as the user moves away from the IRS because the channel vectors $\mathbf{h}_{0,n,k}$, $n=1,\dots, N$, become weaker and  more difficult to estimate.

%\begin{figure*}[t!]
%\begin{mdframed}[linewidth=2pt]
%\begin{center}
%\subfigure[Optimal value of $\tau_{c}$ for an IRS-assisted system to perform as well as a conventional MISO system with $20$ antennas.]{
            %\label{MU1}
            %\includegraphics[width=0.47\textwidth, height=.25\textwidth]{multi_withtau1.eps}
        %} \hspace{.01in}
        %\subfigure[Performance of an  IRS-assisted multi-user MISO system against $N$ under perfect and imperfect CSI.]{
           %\label{MU2}
           %\includegraphics[width=0.47\textwidth, height=.25\textwidth]{multiple_withN1.eps}
        %}
%%\label{singleuserlayout}
%\end{center}
%\caption{Results for a  multi-user system.}
   %\label{MU}
	%\end{mdframed}
%\end{figure*}

\begin{figure*}[!t]
\begin{subfigure}[t]{.48\textwidth}
\tikzset{every picture/.style={scale=.95}, every node/.style={scale=.8}}
\input{multi_withtau1.tex}
\caption{Optimal value of $\tau_{c}$ for an IRS-assisted system to perform as well as a conventional MISO system with $20$ antennas.}
\label{MU1}
\end{subfigure}
\hspace{.05cm}
\begin{subfigure}[t]{.48\textwidth}
\tikzset{every picture/.style={scale=.95}, every node/.style={scale=.8}}
\input{multiple_withN1.tex}
\caption{Performance of an  IRS-assisted multi-user MISO system against $N$ under perfect and imperfect (imp.) CSI.}
\label{MU2}
\end{subfigure}
\caption{Results for a  multi-user system.}
\label{MU}
\end{figure*}
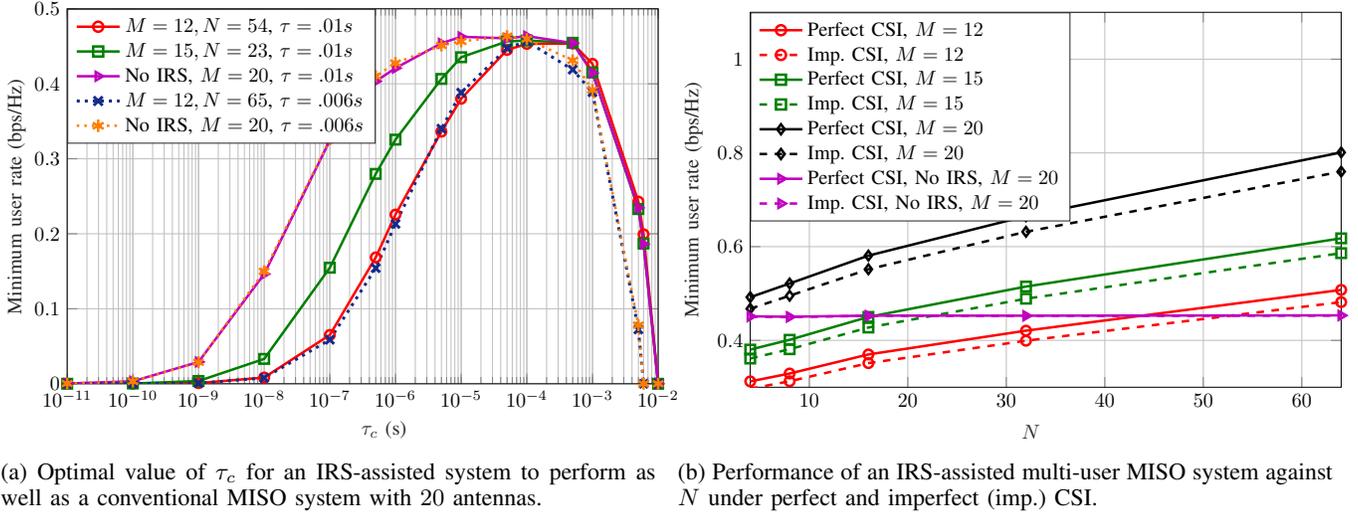

Next we study the minimum user rate performance of multi-user system with the BS placed at $(0,0)$, IRS placed at $(0, 100)$ and users distributed uniformly in the square $(x,y)\in[-30,30]\times[70,130]$ \cite{8741198}. The net achievable rate of user $k$ is given as $\left(1-\frac{\tau_{c}}{\tau} \right) R_{k}$ where $R_{k}=\text{log}_{2}(1+SINR_{k})$, accounting for the rate loss due to channel training. The channel estimation time $\tau_{c}$ needs to be selected optimally to ensure a reasonable quality of channel estimates while minimizing the rate loss due to channel training. In Fig. \ref{MU1} we plot the net achievable minimum rate against $\tau_{c}$ for a conventional MISO system (No IRS) with $20$ active antennas at the BS serving $K=8$ users, assuming a coherence interval of length $10$ms. We also plot the net minimum rates for IRS-assisted systems with a fewer number of active antennas at the BS. The value of $N$ for each $M$ is chosen to ensure that the IRS-assisted system performs as well as the conventional system. The result shows that the IRS-assisted MISO system with $54$ passive reflecting elements at the IRS and only $12$ active antennas at the BS can achieve the same performance as the considered large MISO system.  The same performance can also be achieved with $M=15$ antennas using $N=23$ reflecting elements at the IRS. Therefore, an IRS-assisted MISO system can be an energy-efficient alternative to technologies like massive MISO and network densification. 

 %If the quality of estimates of the IRS-assisted channels could be improved, then the IRS-assisted system could achieve the performance of a conventional system with a much higher number of antennas using smaller sizes of IRS arrays., whereas those with $M=12$ and $M=15$ antennas can achieve the same performance using $N=25$ and $N=11$ reflecting elements respectively

At $2.5$ GHz, the coherence interval $\tau$ ranges from $80$ms for almost stationary scenario (0.5 miles/h users' movement) to $6$ms for dynamic scenario (6 miles/h running speed) \cite{ant}. The solid lines in Fig. \ref{MU1} are plotted for $\tau=.01$s and we observe that the net minimum rate is a unimodal function of $\tau_{c}$. For all considered IRS-assisted MISO systems $\tau_{c}=5\times 10^{-4}$ s is optimal, whereas the conventional system could perform as well with a smaller $\tau_{c}= 10^{-5}$s. When a more dynamic scenario ($\tau=.006$s) is considered in the dotted lines in Fig. \ref{MU1}, the optimal channel estimation time is $\tau_c= 10^{-4}$s, resulting in poorer quality estimates. A higher value of $N$ is then needed to perform as well as the conventional large MISO system. Therefore, the IRS-assisted system is more sensitive to the channel estimation process than the conventional system since a higher number of channels needs to be estimated. The quality of estimates further deteriorates in dynamic scenarios, which results in rate loss. The IRS-assisted system should be able to adapt the channel training time as per the users' speeds. In fact, an important future research direction is to make the channel estimation protocol robust in high speed environments.

Fig. \ref{MU2} corroborates the results of Fig. \ref{MU1} for $\tau=.01$s, by showing the values of $N$ needed to achieve the large MISO gains (yielded by $20$ BS antennas) by using a smaller number of active antennas at the BS and relying on the IRS. We also show the gains of introducing an IRS in a system being served by a BS with $20$ antennas (black curve). With $64$ elements, the IRS can almost double the performance of the conventional system. 

%Note since a larger $\tau_c$ is needed for IRS-assisted system to perform as well as the conventional massive MIMO system, so there is more rate loss in channel training and there is a larger gap between perfect and imperfect CSI curves as compared to the conventional system.

The gains yielded by the IRS have been validated for a sub-6 GHz communication system in this section. However, future networks  will operate at the mmWave frequencies  (30-100 GHz), which are highly susceptible to severe signal attenuation from large structures like buildings, that can result in signal blockages. Introducing IRSs on these structures can enable communication to users with weak or blocked LoS paths by forming strong NLoS paths instead. Therefore, the IRSs will not only yield performance gains similar to what we have seen in this section, but will actually enable communication to users who do not have the direct channel $\mathbf{h}_{d,k}$ in (\ref{ch1}) available.

%We also notice that under channel estimation errors, larger array sizes at the IRS are needed to achieve the same performance as the conventional MISO system with $20$ antennas. For example, under perfect CSI, an IRS-assisted system with $M=10$ antennas can achieve the same performance using only $N=55$  instead of $N=110$ reflecting elements. Moreover, as the value of $N$ increases the performance gap between perfect and imperfect CSI cases for the IRS-assisted system increases significantly since the number of channels to be estimated increases linearly in $N$. Therefore, accurate CSI acquisition is a critical issue in IRS-assisted communication systems. 

\vspace{-.1in}

\section{Further Research Directions}

As IRS-assisted wireless communication is a new paradigm, there exist a number of interesting research directions that can be explored. We summarize them as follows.

\subsection{Modeling Other IRS Implementations}

This work has considered the use of antenna arrays to implement the IRS, where each antenna independently introduces a phase shift onto the impinging EM waves. Recent papers \cite{SRE, LIS_mag1,LISA} envision the concept of SRE enabled by reconfigurable meta-surfaces that have the ability to fully customize the incident EM waves by providing many more functions in addition to the perfect reflection operation considered in this work \cite{SRE}.  The communication-theoretic performance analysis of the SREs enabled by reconfigurable metasurfaces is an important research direction that should utilize accurate communication models that describe the response of the metasurface as a function of its EM properties. In fact, the response of a metasurface will depend on several factors, like its size, the specific material it is made of, as well as on properties of the incident waves, like their angles and the polarization. EM compliant communication models for these surfaces are needed to avoid studying too simplistic and unrealistic system models. Experimentally-validated channel models that provide  path-loss and fast-fading parameters for IRS-assisted links under different implementations are also needed.

%Currently i.i.d. Rayleigh or Rician fading is considered to model the IRS-to-users channels whereas rank-one LoS channel is often considered between the BS and the IRS. The former is not realistic given MIMO channels are known to be almost always correlated in practice whereas the latter constitutes a pinhole channel and limits the capability of the system to efficiently serve more than one user. It will be worth studying correlated channel models for the IRS-to-users links and high rank LoS channel for the BS-to-IRS link.
%
%Extending the preliminary works by replacing the conventional i.i.d. Rayleigh model with a correlated Rayleigh model, would require a spatial correlation model for the IRS.  The conventional statistical models for arrays of discrete antennas are not directly applicable, since IRS is realized using a completely different technology (reconfigurable meta-surfaces or reflect arrays). The correct modeling of the spatial correlation and channels that utilize the inherent correlation structure of the IRS, require significant attention from researchers who are conversant in both communication and electromagnetic theory.  
\vspace{-.05in}
\subsection{CSI Estimation for Large IRSs}
In the current works, the channel sensing limitations of the IRS are ignored and perfect CSI is assumed to be available at the IRS to design its parameters \cite{8741198 , LIS_jour, LISA}. This paper proposed a channel estimation protocol where the channels through the IRS are estimated at the BS by training all the IRS elements, one-by-one. This can yield a large channel training overhead and require longer channel estimation times for large IRSs. 

%However the size of these codebooks will again depend on the size of the IRS.

Alternate channel estimation protocols need to be developed that do not require explicit channel estimation with respect to all the IRS elements. One way to do this is to use beam training to select the IRS phase matrix from quantized codebooks.  Another solution is to embed a few low-power sensors in the IRS, that sense the channels and report them to the BS through the IRS micro-controller. However, this solution will compromise on the almost passive nature of the IRS. 

Moreover, the channel estimation methods need to be adapted based on the technology being used to realize the IRS. For example: reflect-arrays do not require the phase introduced by each of its element to be tuned separately and therefore the proposed protocol is not optimal. The capability of full impinging wavefront sensing provided by reconfigurable metasurfaces also changes the way channel estimation needs to be performed for this implementation of the IRS. 

\vspace{-.05in}
\subsection{Theoretical Performance Limits} 
Almost all existing works \cite{8741198 , LIS_jour, LISA} consider the optimization of the IRS's induced phases by solving complicated, non-convex optimization problems. The optimal phases and the resulting performance are therefore computed numerically and do not have any channel-dependent theoretical expressions (except for the single user, single IRS case in \cite{LISA}) that yield insights into the impact of the channel as well as the radio parameters  on the performance of the IRS-assisted communication systems. Developing tractable analytical frameworks to study the performance of wireless networks enabled by multiple IRSs is an important research direction.

%\vspace{-.03in}
%\subsection{Distributed IRSs}
%Since the IRS is envisioned to be deployed within the LoS of the BS, the BS-to-IRS channel will most likely face the rank deficiency problem. One way to guarantee high rank channels is to utilize distributed IRSs between the BS and the users. Each user will experience a channel that is generated as the sum of as many rank-one channels as the number of IRSs in the system. There is no current work that deals with the use of multiple IRSs, but this would open several research directions, including finding the optimal positions of the IRSs, jointly designing the parameters of all the IRSs and so on.

%The use of distributed LISs with optimum positioning should also be studied. , which might be alleviated to some extent by increasing the inter-element spacing or introducing some deterministic scattering around the IRS
\vspace{-.05in}
\subsection{Integration of IRSs with Existing/Emerging Technologies}

IRSs promise to improve the performance of wireless communication systems by optimizing the environment as compared to the existing solutions that optimize the communication end-points. The integration of IRSs with other existing and emerging technologies, like massive MIMO, mmWave and terahertz communication, small cells, relays, optical communication, vehicular communication etc., needs significant research attention. There are many important directions related to this integration, e.g. the judicious deployment of the IRSs in existing systems, acquiring CSI at the pace of the highly time-varying channels that constitute the vehicular communication systems, optimal user-association schemes for systems with many small cells, distributed IRSs and relays.

%\subsection{IRS-Assisted Vehicular Communication}
%With increasing focus on high mobility communication scenarios, it is important to look into the incorporation of IRSs into vehicular communication systems, where the IRSs can not only improve the coverage area of the underlying cellular network but also help in avoiding signal deterioration and outages when the users move into signal blockage areas. However, these systems entail that the BS attains the CSI and the IRS adapts the induced phases at the pace of the highly time-varying channels. The study of IRS-assisted vehicular communication systems would also require the development of relevant vehicular propagation and channel models.   
%

%\vspace{-.03in}
%\subsection{Hardware Limitations}
%In practice, the performance of LIS-assisted MIMO systems  degrades due to the phase errors introduced by the LIS elements. Also, in practice only low-resolution phase shifters are available instead of the infinite resolution ones considered in most works. The design of LIS having passive reflecting elements with hardware imperfections needs research attention.

\section{Conclusion}
\vspace{-.02in}

In this paper, IRS-assisted wireless communication is envisioned to be an important energy-efficient paradigm  for beyond 5G networks, achieving massive MISO like gains with a much lower number of active antennas at the BS. The  passive antennas constituting the IRS smartly re-configure the signal propagation by introducing phase shifts onto the impinging electromagnetic waves. This paper discussed the evolution of reflective arrays to the IRS concept, outlined the communication model of an IRS-assisted multi-user MISO system and explained how it differentiates from conventional multi-antenna communication models. Later we proposed an MMSE based channel estimation protocol to estimate the IRS-assisted links. We presented performance evaluation results at $2.5$ GHz operating frequency, which  confirmed the superior performance of the proposed system while highlighting its high sensitivity to the quality of the channel estimates.

 %To promote and accelerate the related research activities, this paper outlined the existing challenges and the potential research directions for future works.

\vspace{-.06in}
\bibliographystyle{IEEEtran}
\bibliography{bib}

\end{document}

%% file: singleuserwithd.tex
% This file was created by matlab2tikz.
%
%The latest updates can be retrieved from
%  http://www.mathworks.com/matlabcentral/fileexchange/22022-matlab2tikz-matlab2tikz
%where you can also make suggestions and rate matlab2tikz.
%
\definecolor{mycolor1}{rgb}{0.00000,0.49804,0.00000}%
\definecolor{mycolor2}{rgb}{0.74902,0.00000,0.74902}%
\definecolor{mycolor3}{rgb}{0.85098,0.32549,0.09804}%
\begin{tikzpicture}

\begin{axis}[%
width=\textwidth,
height=.6\textwidth,
scale only axis,
xmin=5,
xmax=120,
xlabel style={font=\normalsize\color{white!15!black}},
xlabel={$d_u$},
ymin=-8,
ymax=45,
ylabel style={at={(axis cs: 10,19)},font=\normalsize\color{white!15!black}},
ylabel={Received SNR (dB)},
axis background/.style={fill=white},
xmajorgrids,
ymajorgrids,
legend style={at={(axis cs: 120,45)},anchor=north east,legend cell align=left,align=left,draw=white!15!black, /tikz/column 2/.style={
                column sep=5pt,
            }}
]
\addplot [color=mycolor1, line width=1.0pt, mark size=2.0pt, mark=o, mark options={solid, mycolor1}]
  table[row sep=crcr]{%
5	36.5096030867947\\
10	32.9476192929858\\
15	28.8452195056797\\
20	25.8687177248412\\
25	23.5801865617676\\
30	22.4432889796191\\
35	22.1860780807461\\
40	23.8147191465741\\
45	26.2369429805308\\
50	27.9771491292307\\
55	26.8240874058721\\
60	23.5218429669802\\
65	20.0488411724108\\
70	17.2108075869963\\
75	14.799530674746\\
80	12.7450129412599\\
85	11.0517558953573\\
90	9.52723350750717\\
95	8.18750079413955\\
100	6.72779962706388\\
105	5.89681617213533\\
110	4.75413605878846\\
115	3.93842801171646\\
120	2.82564068866174\\
};
\addlegendentry{Perfect CSI}

\addplot [color=mycolor2, line width=1.0pt, mark size=2pt, mark=diamond, mark options={solid, mycolor2}]
  table[row sep=crcr]{%
5	36.5053797556059\\
10	32.9410006692009\\
15	28.8321062949702\\
20	25.8470864792254\\
25	23.5508569938823\\
30	22.4155042728421\\
35	22.1631819308555\\
40	23.8018950295539\\
45	26.2309956833287\\
50	27.9733148443201\\
55	26.8193047400879\\
60	23.5102977888556\\
65	20.0217576950824\\
70	17.1576815023079\\
75	14.6953540892426\\
80	12.5746533418974\\
85	10.7943892895429\\
90	9.15121356307064\\
95	7.67215056420433\\
100	6.07343641252331\\
105	5.09618607489345\\
110	3.78403746710252\\
115	2.81886169242843\\
120	1.5230480188006\\
};
\addlegendentry{Imperfect CSI, $\sigma^{2}=-180dBm$}

\addplot [color=red, line width=1.0pt, mark size=2pt, mark=triangle, mark options={solid, rotate=270, red}]
  table[row sep=crcr]{%
5	36.628487220485\\
10	32.6567576111159\\
15	28.8902895875437\\
20	25.8443981692458\\
25	23.6879070313341\\
30	22.1286835855013\\
35	22.157792992309\\
40	23.6322839924353\\
45	26.1921318346195\\
50	27.9133725114739\\
55	26.6766953257836\\
60	23.4638703859429\\
65	19.9327183676036\\
70	16.7871524375022\\
75	14.1222832646228\\
80	11.4658791678493\\
85	9.32993095831599\\
90	7.39481878550063\\
95	5.63042976828952\\
100	4.26243107789258\\
105	3.21752753547135\\
110	1.85871705710602\\
115	0.781483342698705\\
120	-0.207645940253609\\
};
\addlegendentry{Imperfect CSI, $\sigma^2=-170dBm$}

\addplot [color=mycolor3, line width=1.0pt, mark size=2pt, mark=o, mark options={solid, mycolor3}]
  table[row sep=crcr]{%
5	36.7284233482508\\
10	32.7962867211893\\
15	29.4407419957431\\
20	27.1032603879139\\
25	25.0859057071572\\
30	24.6732565664973\\
35	26.0234425330205\\
40	28.2801212912907\\
45	31.3220346197112\\
50	33.4514360782814\\
55	32.1743626339105\\
60	28.7109703104723\\
65	25.1184343638259\\
70	22.0575472184456\\
75	19.3483173343191\\
80	17.1086472801506\\
85	15.1411289779733\\
90	13.2799323644546\\
95	11.8435283688465\\
100	10.3943957088413\\
105	9.13435804556736\\
110	7.85503843876335\\
115	7.11805050714588\\
120	5.99877922566458\\
};
\addlegendentry{Perfect CSI, $N=70$}

\addplot [color=mycolor1, dashed, line width=1.0pt, mark size=2.2pt, mark=o, mark options={solid, mycolor1}]
  table[row sep=crcr]{%
5	36.4775860887426\\
10	32.8837043388762\\
15	28.6533324635528\\
20	25.3508926042585\\
25	22.3488712195812\\
30	19.8646952190884\\
35	17.4118336832661\\
40	15.7689682857676\\
45	14.13404709634\\
50	12.0818599205242\\
55	11.0746921337067\\
60	9.43704321828825\\
65	8.17568207691817\\
70	6.78388513869871\\
75	5.82678970357168\\
80	4.98305009178071\\
85	4.14159460768953\\
90	3.30207719259379\\
95	2.52422328199468\\
100	1.16482409582234\\
105	0.852004592309049\\
110	-0.0520858331314162\\
115	-0.639015618624853\\
120	-1.62401556432476\\
};
%\addlegendentry{Perfect CSI, No IRS}

\addplot [color=mycolor2, dashed, line width=1.0pt, mark size=2.4pt, mark=diamond, mark options={solid, mycolor2}]
  table[row sep=crcr]{%
5	36.4775860736427\\
10	32.8837043029955\\
15	28.653332379536\\
20	25.3508924115687\\
25	22.3488708483246\\
30	19.8646945306723\\
35	17.4118325201776\\
40	15.7689665646224\\
45	14.1340446440571\\
50	12.081856023128\\
55	11.0746871590264\\
60	9.43703608567791\\
65	8.17567254052102\\
70	6.78387173334695\\
75	5.8267719959654\\
80	4.98302946105697\\
85	4.14156973982043\\
90	3.30204614085084\\
95	2.52418536949336\\
100	1.16477337901893\\
105	0.851951659146591\\
110	-0.0521535488368301\\
115	-0.639092543349075\\
120	-1.62410890557921\\
};
%\addlegendentry{Imp. CSI, No IRS, $\sigma^2=-180dBm$}

\addplot [color=red, dashed, line width=1.0pt, mark size=1.6pt, mark=triangle, mark options={solid, rotate=270, red}]
  table[row sep=crcr]{%
5	36.6137244116182\\
10	32.6197806075662\\
15	28.7720092329507\\
20	25.4852368008407\\
25	22.7454162533465\\
30	19.8571046278803\\
35	17.7042563158538\\
40	15.6418817083468\\
45	14.1880147549095\\
50	12.3313943504238\\
55	10.8237543696083\\
60	9.18975818283072\\
65	8.29231916560219\\
70	7.04501672820938\\
75	6.03115647893202\\
80	4.96890526703314\\
85	4.06745281068109\\
90	3.09926452591862\\
95	2.15607346423534\\
100	1.47491998978442\\
105	0.987569822814023\\
110	-0.0235127404484687\\
115	-0.712836942471135\\
120	-1.48617136287558\\
};
%\addlegendentry{Imp. CSI, No IRS, $\sigma^{2}=-170dBm$}

\node at (axis cs: 7,5) [anchor = west] {\normalsize Dashed lines: Without IRS};

\end{axis}

\end{tikzpicture}%

%% file: multi_withtau1.tex
% This file was created by matlab2tikz.
%
%The latest updates can be retrieved from
%  http://www.mathworks.com/matlabcentral/fileexchange/22022-matlab2tikz-matlab2tikz
%where you can also make suggestions and rate matlab2tikz.
%
\definecolor{mycolor1}{rgb}{0.00000,0.49804,0.00000}%
\definecolor{mycolor2}{rgb}{0.74902,0.00000,0.74902}%
\definecolor{mycolor3}{rgb}{0.07843,0.16863,0.54902}%
\definecolor{mycolor4}{rgb}{1,0.49,0}%
\begin{tikzpicture}

\begin{axis}[%
width=\textwidth,
height=.635\textwidth,
scale only axis,
xmode=log,
xmin=1e-11,
xmax=0.01,
xminorticks=true,
xlabel style={font=\normalsize\color{white!15!black}},
xlabel={$\tau_c\text{ (s)}$},
ymin=0,
ymax=0.5,
ylabel style={at={(axis cs: 2*1e-11,.25)},font=\normalsize\color{white!15!black}},
ylabel={Minimum user rate (bps/Hz)},
axis background/.style={fill=white},
xmajorgrids,
xminorgrids,
ymajorgrids,
legend style={at={(axis cs: 1e-11,.5)},anchor=north west,legend cell align=left,align=left,draw=white!15!black, /tikz/column 2/.style={
                column sep=5pt,
            }}
]
\addplot [color=red, line width=1.0pt,mark size=2pt, mark=o, mark options={solid, red}]
  table[row sep=crcr]{%
1e-11	8.53065093021781e-06\\
1e-10	8.8134088305189e-05\\
1e-09	0.00087464992580316\\
1e-08	0.00829826659358573\\
1e-07	0.0650967330914321\\
5e-07	0.1685207701899\\
1e-06	0.225647321538493\\
5e-06	0.336195352421226\\
1e-05	0.380107901456232\\
5e-05	0.445223273736918\\
0.0001	0.453229599814671\\
0.0005	0.453695236893268\\
0.001	0.42658842482197\\
0.005	0.242960356294466\\
0.006	0.19929195519173\\
0.01	0\\
};
\addlegendentry{$M=12, N=54$, $\tau=.01s$}

\addplot [color=mycolor1, line width=1.0pt,mark size=2pt, mark=square, mark options={solid, mycolor1}]
  table[row sep=crcr]{%
1e-11	3.71543557284597e-05\\
1e-10	0.000353439220267405\\
1e-09	0.0036620187795927\\
1e-08	0.0331520476355269\\
1e-07	0.154772957564089\\
5e-07	0.279862579534601\\
1e-06	0.325551950404064\\
5e-06	0.406508613327124\\
1e-05	0.435449765755451\\
5e-05	0.456554933130605\\
0.0001	0.457709867250609\\
0.0005	0.454714057641576\\
0.001	0.41517272293928\\
0.005	0.2329561390256\\
0.006	0.187015607212009\\
0.01	0\\
};
\addlegendentry{$M=15, N=23$, $\tau=.01s$}

\addplot [color=mycolor2, line width=1.0pt,mark size=2pt, mark=triangle, mark options={solid, rotate=270, mycolor2}]
  table[row sep=crcr]{%
1e-11	0.000323682339049433\\
1e-10	0.00331071767829098\\
1e-09	0.0291026694560888\\
1e-08	0.146315704929293\\
1e-07	0.324476566417473\\
5e-07	0.403346274440038\\
1e-06	0.420907157002619\\
5e-06	0.454061358062926\\
1e-05	0.462711284933758\\
5e-05	0.46061031733537\\
0.0001	0.463605643788282\\
0.0005	0.454308393976296\\
0.001	0.41447744484808\\
0.005	0.234589391562136\\
0.006	0.186288535999378\\
0.01	0\\
};
\addlegendentry{No IRS, $M=20$, $\tau=.01s$}

\addplot [color=mycolor3, dotted, line width=1.2pt,mark size=2.5pt, mark=x, mark options={solid, mycolor3}]
  table[row sep=crcr]{%
1e-11	7.66976477044628e-06\\
1e-10	8.03049822947655e-05\\
1e-09	0.000760310462526978\\
1e-08	0.00778128760574759\\
1e-07	0.0587296330254907\\
5e-07	0.154171836043115\\
1e-06	0.213159563078753\\
5e-06	0.340180236553249\\
1e-05	0.388234300858417\\
5e-05	0.447452163360104\\
0.0001	0.456527951574685\\
0.0005	0.418859327117412\\
0.001	0.388787068320415\\
0.005	0.0724290180524932\\
0.006	0\\
0.01	0\\
};
\addlegendentry{$M=12, N=65$, $\tau=.006s$}

\addplot [color=mycolor4, dotted, line width=1.0pt,mark size=2.5pt, mark=asterisk, mark options={solid, mycolor4}]
  table[row sep=crcr]{%
1e-11	0.000345531146580018\\
1e-10	0.00323858938141567\\
1e-09	0.0286972412052405\\
1e-08	0.149867704876676\\
1e-07	0.325897114816509\\
5e-07	0.410179854234079\\
1e-06	0.42768466316812\\
5e-06	0.451212981155773\\
1e-05	0.45655171543517\\
5e-05	0.463549301809015\\
0.0001	0.459740102195092\\
0.0005	0.43121436799051\\
0.001	0.391776979432874\\
0.005	0.0781459366673763\\
0.006	0\\
0.01	0\\
};
\addlegendentry{No IRS, $M=20$, $\tau=.006s$}

\end{axis}
\end{tikzpicture}%

%% file: multiple_withN1.tex
% This file was created by matlab2tikz.
%
%The latest updates can be retrieved from
%  http://www.mathworks.com/matlabcentral/fileexchange/22022-matlab2tikz-matlab2tikz
%where you can also make suggestions and rate matlab2tikz.
%
\definecolor{mycolor1}{rgb}{0.00000,0.49804,0.00000}%
\definecolor{mycolor2}{rgb}{0.74902,0.00000,0.74902}%
\begin{tikzpicture}

\begin{axis}[%
width=\textwidth,
height=.635\textwidth,
scale only axis,
xmin=4,
xmax=64,
xlabel style={font=\normalsize\color{white!15!black}},
xlabel={$N$},
ymin=0.3,
ymax=1.1,
ylabel style={at={(axis cs: 5.4,.7)},font=\normalsize\color{white!15!black}},
ylabel={Minimum user rate (bps/Hz)},
axis background/.style={fill=white},
xmajorgrids,
ymajorgrids,
legend style={at={(axis cs: 4,1.1)},anchor=north west,legend cell align=left,align=left,draw=white!15!black, /tikz/column 2/.style={
                column sep=5pt,
            }}
]
\addplot [color=red, line width=1.0pt,mark size=2pt, mark=o, mark options={solid, red}]
  table[row sep=crcr]{%
4	0.312467895276159\\
8	0.329246214766964\\
16	0.369972960306121\\
32	0.420702063761817\\
64	0.50768381681587\\
};
\addlegendentry{Perfect CSI, $M=12$}

\addplot [color=red, dashed, line width=1.0pt,mark size=2pt, mark=o, mark options={solid, red}]
  table[row sep=crcr]{%
4	0.296821786658333\\
8	0.312745461021093\\
16	0.351416363973731\\
32	0.39925328375096\\
64	0.481603105325068\\
};
\addlegendentry{Imp. CSI, $M=12$}

\addplot [color=mycolor1, line width=1.0pt,mark size=2pt, mark=square, mark options={solid, mycolor1}]
  table[row sep=crcr]{%
4	0.380296680457242\\
8	0.401217226842354\\
16	0.450001494809482\\
32	0.514843864803539\\
64	0.61771050784821\\
};
\addlegendentry{Perfect CSI, $M=15$}

\addplot [color=mycolor1, dashed, line width=1.0pt,mark size=2pt, mark=square, mark options={solid, mycolor1}]
  table[row sep=crcr]{%
4	0.361250104612408\\
8	0.381135897473749\\
16	0.427405244918577\\
32	0.488904860260466\\
64	0.586109822500394\\
};
\addlegendentry{Imp. CSI, $M=15$}

\addplot [color=black, line width=1.0pt,mark size=2pt, mark=diamond, mark options={solid, black}]
  table[row sep=crcr]{%
4	0.492098506294189\\
8	0.52135628606306\\
16	0.580805396593054\\
32	0.665005724816241\\
64	0.801023275744232\\
};
\addlegendentry{Perfect CSI, $M=20$}

\addplot [color=black, dashed, line width=1.0pt,mark size=2pt, mark=diamond, mark options={solid, black}]
  table[row sep=crcr]{%
4	0.467467991503537\\
8	0.495242826257553\\
16	0.551671879646892\\
32	0.631664536779068\\
64	0.760104929632738\\
};
\addlegendentry{Imp. CSI, $M=20$}

\addplot [color=mycolor2, line width=1.0pt,mark size=2pt, mark=triangle, mark options={solid, rotate=270, mycolor2}]
  table[row sep=crcr]{%
4	0.451328203329607\\
8	0.450350354083682\\
16	0.452801840174108\\
32	0.452662964280237\\
64	0.453390070014103\\
};
\addlegendentry{Perfect CSI, No IRS, $M=20$}

\addplot [color=mycolor2, dashed, line width=1.0pt,mark size=2pt, mark=triangle, mark options={solid, rotate=270, mycolor2}]
  table[row sep=crcr]{%
4	0.450715939288566\\
8	0.449765045927294\\
16	0.452188342724426\\
32	0.452084166670195\\
64	0.452846455390333\\
};
\addlegendentry{Imp. CSI, No IRS, $M=20$}

\end{axis}
\end{tikzpicture}%